\documentclass[a4paper]{jpconf} 
\usepackage{graphicx}
\usepackage{amssymb,amsmath,cite} 
\def\lamb#1#2{$^{#1}_{\Lambda}${#2}}
\def\lam#1#2{$^{#1}_{~\Lambda}${#2}}

\begin{document}

\title{Charge symmetry breaking in light $\Lambda$ hypernuclei} 
\author{Avraham Gal$^1$ and Daniel Gazda$^{2,3}$} 
\address{$^1$ Racah Institute of Physics, The Hebrew University, 
91904 Jerusalem, Israel \\  
$^2$ Department of Physics, Chalmers University of Technology, SE-412 96 
G\"{o}teborg, Sweden \\ 
$^3$ Nuclear Physics Institute, 25068 \v{R}e\v{z}, Czech Republic} 

\ead{avragal@savion.huji.ac.il} 

\begin{abstract} 
Charge symmetry breaking (CSB) is particularly strong in the $A=4$ mirror 
hypernuclei \lamb{4}{H}--\lamb{4}{He}. Recent four-body no-core shell model 
calculations that confront this CSB by introducing $\Lambda$-$\Sigma^0$ mixing 
to leading-order chiral effective field theory hyperon-nucleon potentials are 
reviewed, and a shell-model approach to CSB in $p$-shell $\Lambda$ hypernuclei 
is outlined. 
\end{abstract}

\section{Introduction} 
\label{sec:intro} 

Charge symmetry of the strong interactions arises in QCD upon neglecting the 
few-MeV mass difference of up and down quarks. With baryon masses of order 
GeV, charge symmetry should break down at the level of 10$^{-3}$ in nuclei. 
The lightest nuclei to exhibit charge symmetry breaking (CSB) are the $A$=3 
mirror nuclei $^3$H--$^3$He, where CSB contributes about 70 keV out of the 
764~keV Coulomb-dominated binding-energy difference. This CSB contribution is 
indeed of order 10$^{-3}$ with respect to the strong interaction contribution 
in realistic $A$=3 binding energy calculations, and is also consistent in both 
sign and size with the scattering-length difference $a_{pp}-a_{nn}\approx 
1.7$~fm~\cite{miller06}. It can be explained by $\rho^0\omega$ mixing in 
one-boson exchange models of the $NN$ interaction, or by considering $N\Delta$ 
intermediate-state mass differences in models limited to pseudoscalar meson 
exchanges \cite{mach01}. In practice, introducing two charge dependent contact 
interaction terms in chiral effective field theory ($\chi$EFT) applications, 
one accounts quantitatively for the charge dependence of the low energy 
$NN$ scattering parameters and, thereby, also for the $A$=3 mirror nuclei 
binding-energy difference~\cite{entem03}. CSB is manifest, of course, also 
in heavier nuclei. 

\begin{figure}[hbt] 
\begin{center} 
\includegraphics[width=0.48\textwidth,height=5cm]{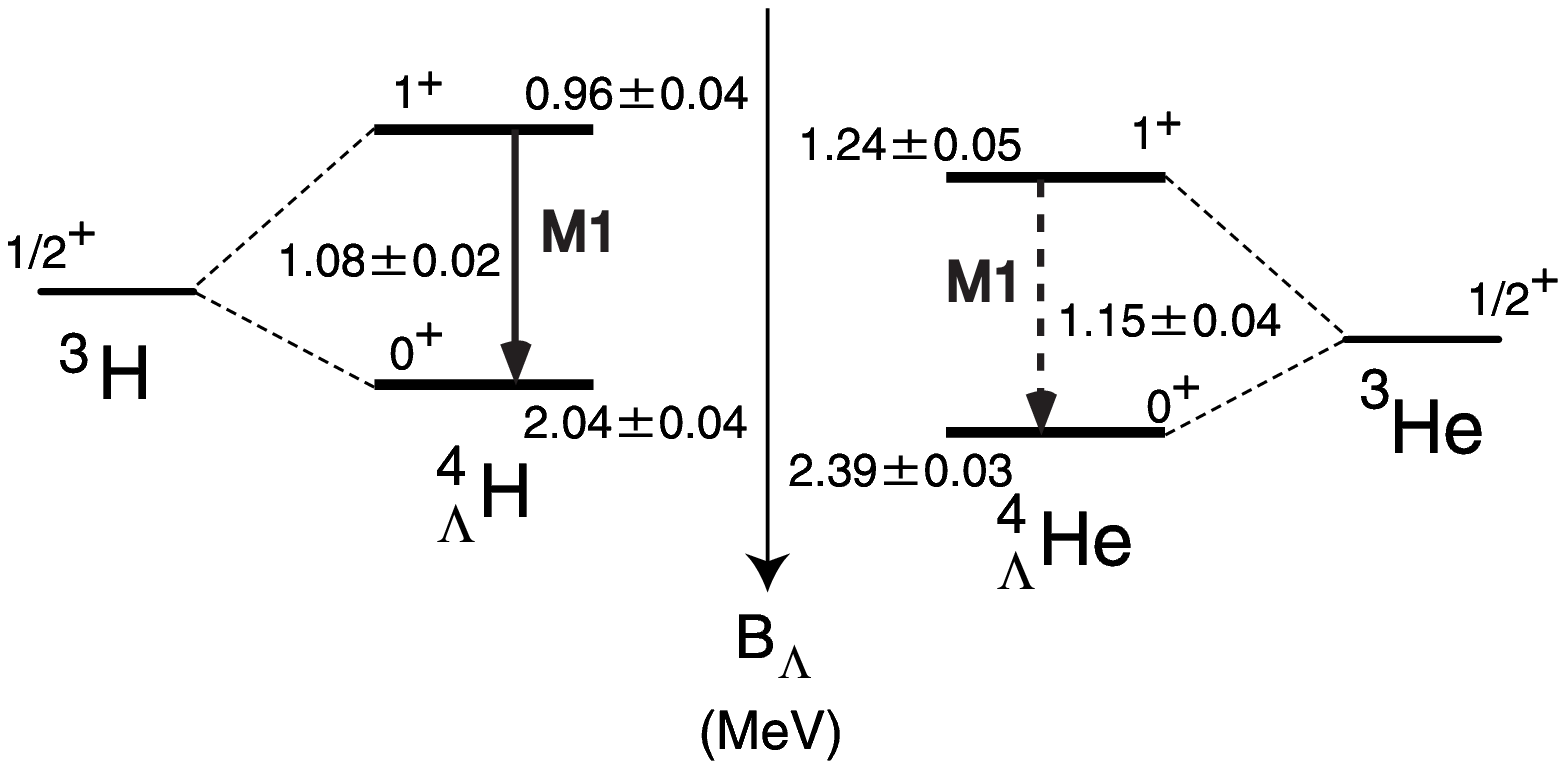} 
\includegraphics[width=0.48\textwidth,height=5cm]{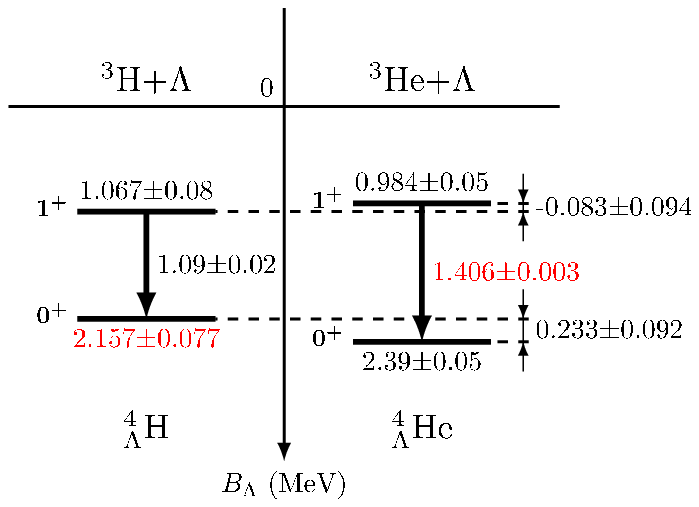} 
\end{center} 
\caption{\lamb{4}{H}-\lamb{4}{He} level diagram, before (left panel) and after 
(right panel) the recent measurements of the \lamb{4}{He} excitation energy 
$E_{\gamma}(1^+_{\rm exc}\to 0^+_{\rm g.s.})$ at J-PARC~\cite{E13}, and of 
the \lamb{4}{H} $0^+_{\rm g.s.}$ binding energy at MAMI~\cite{MAMI15,MAMI16}, 
both highlighted in red in the online version. CSB splittings are shown to 
the very right of the \lamb{4}{He} levels. Figure adapted from~\cite{MAMI16}.} 
\label{fig:A=4} 
\end{figure}

In $\Lambda$ hypernuclei, isospin invariance excludes one pion exchange (OPE) 
from contributing to $\Lambda N$ strong-interaction matrix elements. However, 
it was pointed out by Dalitz and Von Hippel (DvH) that the SU(3) octet 
$\Lambda_{I=0}$ and $\Sigma^0_{I=1}$ hyperons are admixed in the physical 
$\Lambda$ hyperon, thus generating a long-range OPE $\Lambda N$ CSB potential 
$V_{\rm CSB}^{\rm OPE}$~\cite{DvH64}. For the mirror \lamb{4}{H}--\lamb{4}{He} 
ground-state (g.s.) levels built on the $^3$H--$^3$He g.s. cores, and using 
the DvH purely central wavefunction, the OPE CSB contribution amounts to 
$\Delta B_{\Lambda}^{J=0}$$\approx$95~keV where $\Delta B_{\Lambda}^J\equiv 
B_{\Lambda}^J$(\lamb{4}{He})$-$$B_{\Lambda}^J$(\lamb{4}{H}). This is also 
confirmed in our present calculations in which tensor contributions add up 
$\approx$100~keV. Shorter-range CSB meson-mixing contributions appear to 
be much smaller~\cite{coon99}. Remarkably, the OPE overall contribution 
of $\approx$200~keV to the CSB splitting of the \lamb{4}{H}--\lamb{4}{He} 
mirror g.s. levels roughly agrees with the large observed g.s. CSB splitting 
$\Delta B^{J=0}_{\Lambda}$=233$\pm$92~keV shown in Fig.~\ref{fig:A=4} 
which is of order 10$^{-2}$ with respect to the $\Lambda$ nuclear strong 
interaction contribution in realistic binding energy calculations of the 
$A$=4 hypernuclei. Hence, CSB in $\Lambda$ hypernuclei is likely to be 
almost one order of magnitude stronger than in ordinary nuclei. 

\begin{figure}[hbt] 
\begin{center} 
\includegraphics[width=0.5\textwidth]{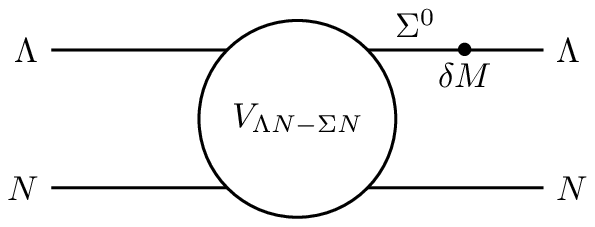} 
\end{center} 
\caption{CSB $\Lambda N$ interaction diagram describing a CS $V_{\Lambda N - 
\Sigma N}$ interaction followed by a CSB $\Lambda-\Sigma^0$ mass-mixing 
vertex.} 
\label{fig:mass} 
\end{figure}

In addition to OPE, $\Lambda-\Sigma^0$ mixing affects also shorter range 
meson exchanges (e.g. $\rho$) that in $\chi$EFT are replaced by contact 
terms. Quite generally, in baryon-baryon models that include {\it explicitly} 
a charge-symmetric (CS) $\Lambda N\leftrightarrow\Sigma N$ ($\Lambda\Sigma$) 
coupling, the direct $\Lambda N$ matrix element of $V_{\rm CSB}$ is obtained 
from a strong-interaction CS $\Lambda\Sigma$ coupling matrix element 
$\langle N\Sigma|V_{\rm CS}|N\Lambda\rangle$ by 
\begin{equation} 
\langle N\Lambda|V_{\rm CSB}|N\Lambda\rangle = -0.0297\,\tau_{Nz}\,\frac{1}
{\sqrt{3}}\,\langle N\Sigma|V_{\rm CS}|N\Lambda\rangle , 
\label{eq:OME} 
\end{equation} 
where the $z$ component of the nucleon isospin Pauli matrix ${\vec\tau}_N$ 
assumes the values $\tau_{Nz}=\pm 1$ for protons and neutrons, respectively, 
the isospin Clebsch-Gordan coefficient $1/\sqrt{3}$ accounts for the 
$N\Sigma^0$ amplitude in the $I_{NY}=1/2$ $N\Sigma$ state, and the space-spin 
structure of this $N\Sigma$ state is taken identical to that of the $N\Lambda$ 
state sandwiching $V_{\rm CSB}$. The 3\% CSB scale factor $-$0.0297 in 
Eq.~(\ref{eq:OME}) follows by evaluating the $\Lambda-\Sigma^0$ mass mixing 
matrix element $\langle\Sigma^0|\delta M|\Lambda\rangle$ from SU(3) mass 
formulae \cite{DvH64,gal15}. The corresponding diagram for generating $\langle 
N\Lambda|V_{\rm CSB}|N\Lambda\rangle$ is shown in Fig.~\ref{fig:mass}, 
demonstrating explicitly the $\delta M$ CSB insertion.  

Since the CSB $\Lambda N$ matrix element in Eq.~(\ref{eq:OME}) is given in 
terms of strong-interaction CS $\Lambda\Sigma$ coupling, one wonders how 
strong the latter is in realistic microscopic $YN$ interaction models. 
Recent four-body calculations of $_{\Lambda}^4{\rm He}$ levels \cite{gazda14}, 
using the Bonn-J\"{u}lich leading order~(LO) $\chi$EFT $YN$ CS potential 
model~\cite{LO06}, show that almost 40\% of the $0^+_{\rm g.s.}\to 1^+_{\rm 
exc}$ excitation energy $E_x$ arises from $\Lambda\Sigma$ coupling. This also 
occurs in the NSC97 models~\cite{NSC97} as demonstrated by Akaishi 
{\it et al}~\cite{akaishi00}. With $\Lambda\Sigma$ matrix elements of order 
10~MeV, the 3\% CSB scale factor in Eq.~(\ref{eq:OME}) suggests a CSB splitting 
$\Delta E_x \sim 300$~keV, in good agreement with the observed splitting 
$E_{\rm x}$(\lamb{4}{He})$-E_{\rm x}$(\lamb{4}{H})$=320\pm 20$~keV~\cite{E13}, 
see Fig.~\ref{fig:A=4} (right) which also shows a relatively large splitting 
of the $A$=4 mirror hypernuclear g.s. levels, $\Delta B^{J=0}_{\Lambda}$=233$
\pm$92~keV~\cite{MAMI15,MAMI16}, with respect to the $\approx$70~keV CSB 
splitting in the mirror core nuclei $^3$H and $^3$He. 

Here we review recent {\it ab-initio} no-core shell model (NCSM) calculaions 
of the $A$=4 $\Lambda$ hypernuclei~\cite{gg15,gg16} using a LO $\chi$EFT 
$YN$ CS interaction model~\cite{LO06} in which CSB is generated by 
implementing Eq.~(\ref{eq:OME}). We also briefly review a shell model 
approach~\cite{gal15}, confronting it with some available data in the 
$p$ shell.

\section{LO $\chi$EFT $YN$ interactions} 
\label{sec:YN} 

N3LO $NN$ \cite{entem03} and N2LO $NNN$ interactions \cite{navratil07}, 
both with momentum cutoff $\Lambda=500$~MeV, are used in our calculations. 
For hyperons, the Bonn-J\"{u}lich SU(3)-based LO $YN$ interaction is used, 
plus $V_{\rm CSB}$ evaluated from it according to Eq.~(\ref{eq:OME}). 
At LO, $V_{YN}$ consists of regularized pseudoscalar (PS) $\pi$, $K$ and 
$\eta$ meson exchanges with coupling constants constrained by SU(3)$_{\rm f}$, 
plus five central interaction contact terms simulating the short range 
behavior of the $YN$ coupled channel interactions, all of which are 
regularized with a cutoff momentum $\Lambda\geq m_{\rm PS}$, varied from 550 
to 700 MeV. Two of the five contact terms connect $\Lambda N$ to $\Sigma N$ in 
spin-singlet and triplet $s$-wave channels, and are of special importance for 
the calculation of CSB splittings. The dominant meson exchange interaction is 
OPE which couples the $\Lambda N$ channel exclusively to the $I=\frac{1}{2}$ 
$\Sigma N$ channel. $K$-meson exchange also couples these two $YN$ channels. 
This $V_{YN}^{\rm LO}$ reproduces reasonably well, with $\chi^2/(\rm{d.o.f.})
\approx 1$, the scarce $YN$ low-energy scattering data. It also reproduces 
the binding energy of \lamb{3}{H}, with a calculated value $B_{\Lambda}
$(\lamb{3}{H})=110$\pm$10~keV for $\Lambda$=600~MeV \cite{wirth14}, consistent 
with experiment (130$\pm$50~keV \cite{davis05}) and with Faddeev calculations 
reported by Haidenbauer {\it et al}~\cite{nogga07}. Isospin conserving matrix 
elements of $V_{YN}^{\rm LO}$ are evaluated in a momentum-space particle basis 
accounting for mass differences within baryon iso-multiplets, while isospin 
breaking $(I_{NN}=0)\leftrightarrow (I_{NN}=1)$ and $(I_{YN}=\frac{1}{2})
\leftrightarrow (I_{YN}=\frac{3}{2})$ transitions are suppressed. 
The Coulomb interaction between charged baryons is included.

\section{NCSM hypernuclear calculations} 
\label{sec:NCSM} 

The NCSM approach to few-body calculations uses translationally invariant 
harmonic-oscillator (HO) bases expressed in terms of relative Jacobi 
coordinates~\cite{navratil00} in which two-body and three-body interaction 
matrix elements are evaluated. Antisymmetrization is imposed with respect to 
nucleons, and the resulting Hamiltonian is diagonalized in a finite HO basis, 
admitting all HO excitation energies $N\hbar\omega$, $N\leq N_{\rm max}$, 
up to $N_{\rm max}$ HO quanta. This NCSM nuclear technique was extended 
recently to light hypernuclei \cite{gazda14,wirth14}. While it was 
possible to obtain fully converged binding energies, with keV precision 
for the $A$=3 core nuclei $^3$H and $^3$He, it was not computationally 
feasible to perform calculations with sufficiently large $N_\text{max}$ 
to demonstrate convergence for \lamb{4}{H} and \lamb{4}{He}. In these cases 
extrapolation to an infinite model space, $N_\text{max}\rightarrow\infty$, 
had to be employed. For details see Ref.~\cite{gg16}. We note that $\Delta 
B_{\Lambda}$, and to a lesser extent $B_{\Lambda}$, exhibit fairly weak 
$N_\text{max}$ and $\omega$ dependence compared to the behavior of absolute 
energies, and the employed extrapolation scheme was found sufficiently robust. 
While normally using $N_\text{max}\to\infty$ extrapolated values based on the 
last three $N_\text{max}$ values, it was found that including the last four 
$N_\text{max}$ values in the fit resulted in $\Delta B_\Lambda$ values that 
differed by $\lesssim 10$~keV. 

Calculations consisting of fully converged $A$=3 core binding energies 
(8.482 MeV for $^3$H and 7.720 MeV for $^3$He) and (\lamb{4}{H}, \lamb{4}{He}) 
$0^+_{\rm g.s.}$ and $1^{+}_{\rm exc}$ binding energies extrapolated to 
infinite model spaces from $N_{\rm max}=18(14)$ for $J=0(1)$ are reported here. 
The $NNN$ interaction, was excluded from most of the hypernuclear calculations 
after verifying that, in spite of adding almost 80 keV to the $\Lambda$ 
separation energies $B^{J=0}_{\Lambda}$ and somewhat less to $B^{J=1}_{
\Lambda}$, its inclusion makes a difference of only a few keV for the CSB 
splittings $\Delta B^{J}_{\Lambda}$ in both the $0^+_{\rm g.s.}$ and $1^+_{
\rm exc}$ states. 

\begin{table}[htb] 
\caption{CS averages (in MeV) of $B^{J}_{\Lambda}$(\lamb{4}{H}) and 
$B^{J}_{\Lambda}$(\lamb{4}{He}) in four-body calculations using LO $\chi$EFT 
$YN$ \cite{LO06} and NLO $\chi$EFT $YN$ \cite{NLO13} interaction models.} 
\begin{center} 
\begin{tabular}{ccccc} 
\br 
 & LO (present) & LO~~\cite{nogga13} & NLO~~\cite{nogga13} & 
Exp.~(Fig.~\ref{fig:A=4}) \\ 
\mr 
$B^{J=0}_{\Lambda}$ & 2.37$^{+0.20}_{-0.13}$ & 2.5$\pm$0.1 & 
1.53$^{+0.08}_{-0.06}$ & 2.27$\pm$0.09 \\
$B^{J=1}_{\Lambda}$ & 1.08$^{+0.58}_{-0.47}$ & 1.4$^{+0.5}_{-0.4}$ & 
0.83$^{+0.07}_{-0.10}$ & 1.03$\pm$0.09 \\ 
$E_{\rm x}$(0$^+_{\rm g.s.}$$\to$1$^+_{\rm exc}$) & 
1.29$\pm$0.38 & 1.05$\pm$0.25 & 0.71$\pm$0.04 & 1.25$\pm$0.02 \\ 
\br 
\end{tabular} 
\label{tab:nogga} 
\end{center} 
\end{table}

Table~\ref{tab:nogga} lists results obtained for the $A$=4 hypernuclear levels 
in the present LO-$YN$ NCSM calculation with $V_{\rm CSB}$, and in Nogga's 
\cite{nogga13} LO- and NLO-$YN$ Faddeev-Yakubovsky calculations without 
$V_{\rm CSB}$. To provide meaningful comparison, the `present' column lists 
CS averages over mirror levels in \lamb{4}{H} and \lamb{4}{He}. The two 
LO columns are consistent with each other within the cited uncertainties, 
which are particularly large for $J=1$, and both agree with experiment within 
these uncertainties. Uncertainties reflect the resulting cutoff dependence in 
the chosen $\Lambda$ range. The NLO results are almost $\Lambda$ independent, 
as inferred from their small uncertainties. However, NLO disagrees strongly 
with experiment, particularly for $J=0$ and for the accurately determined 
$E_x$. It would be interesting in future work to modify the existing NLO 
$\chi$EFT version~\cite{NLO13,nogga13} by refitting the $\Lambda\Sigma$ 
contact terms to both $B^{J=0,1}_{\Lambda}$(A=4) CS-averaged values, and then 
apply the CSB generating equation (\ref{eq:OME}) in four-body calculations of 
\lamb{4}{H}--\lamb{4}{He}.

\section{CSB in $s$-shell hypernuclei}
\label{sec:s} 

\begin{figure}[hbt] 
\begin{center} 
\includegraphics[width=0.48\textwidth,height=5cm]{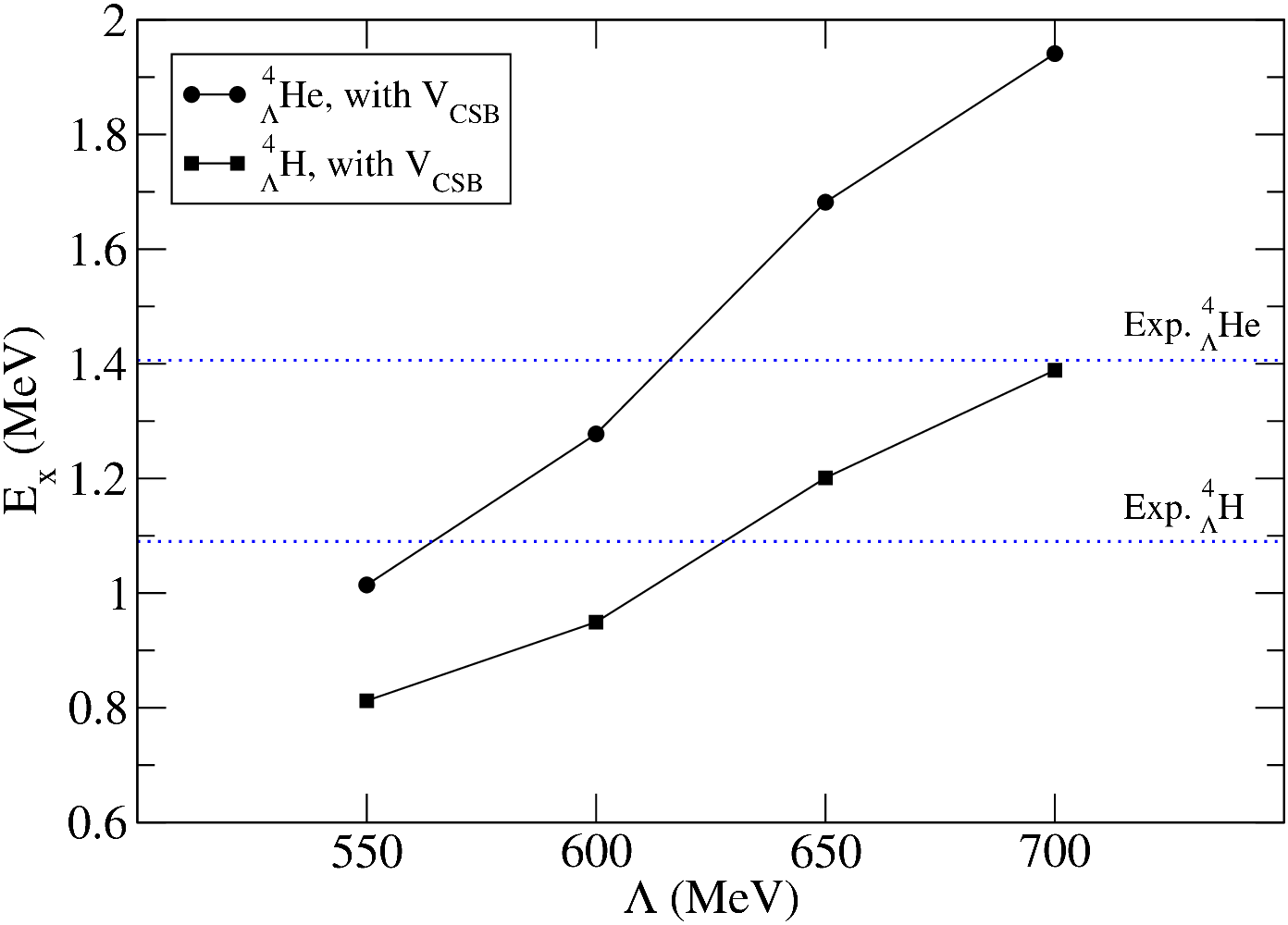} 
\includegraphics[width=0.48\textwidth,height=5cm]{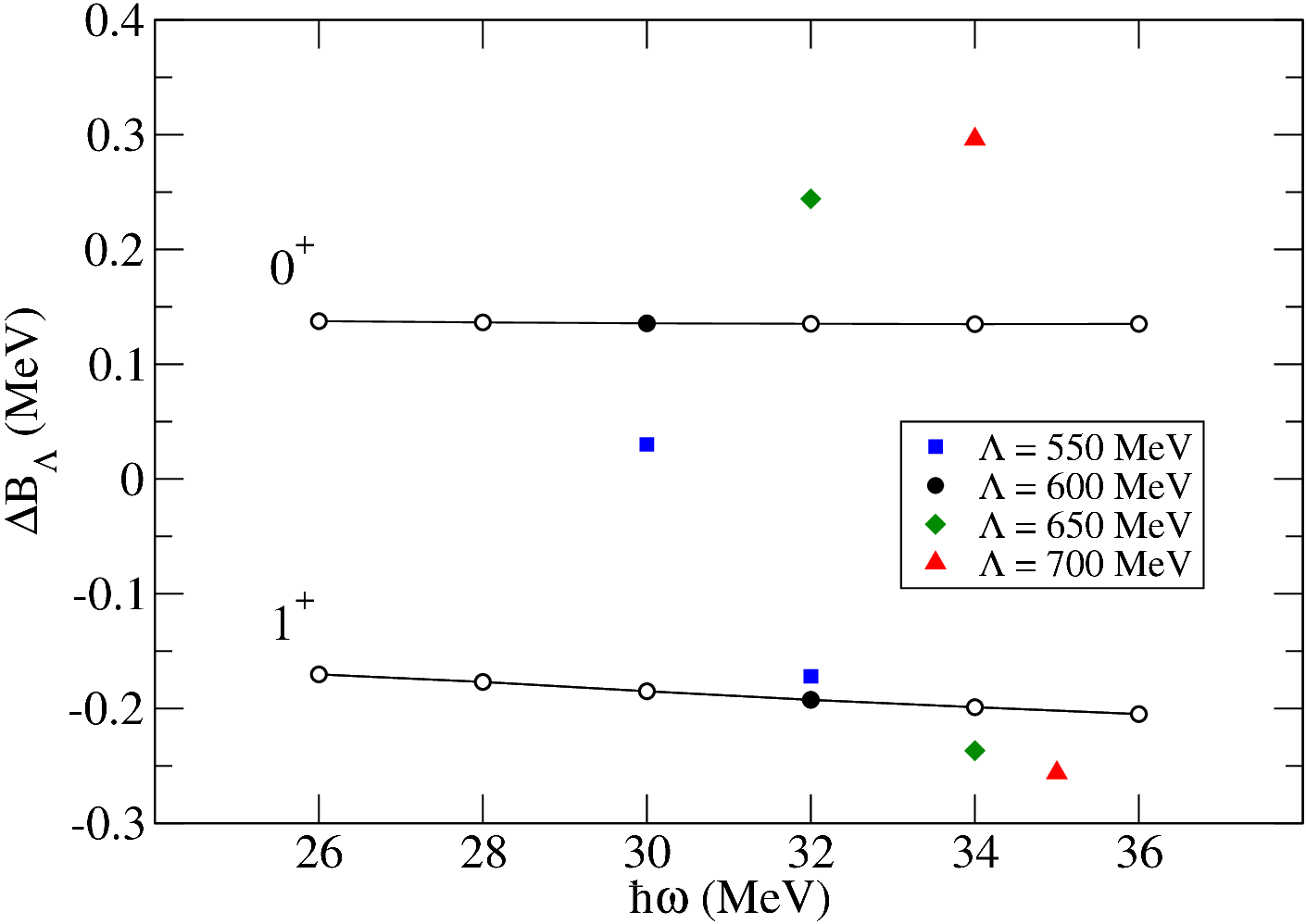} 
\end{center} 
\caption{NCSM calculations of \lamb{4}{H} and \lamb{4}{He}, using CS LO 
$\chi$EFT $YN$ interactions~\cite{LO06} and $V_{\rm CSB}$, Eq.~(\ref{eq:OME}), 
derived from these CS interactions. Left: momentum cutoff dependence of 
excitation energies $E_{\rm x}$(0$^+_{\rm g.s.}$$\to$1$^+_{\rm exc}$). 
The $\gamma$-ray measured values of $E_{\rm x}$ from Fig.~\ref{fig:A=4} are 
marked by dotted horizontal lines. Right: HO $\hbar\omega$ dependence, for 
$\Lambda$=600~MeV, of the separation-energy differences $\Delta B_{\Lambda}^J$ 
for $0^+_{\rm g.s.}$ (upper curve) and for $1^+_{\rm exc}$ (lower curve). 
Results for other values of $\Lambda$ are shown at the respective absolute 
variational energy minima. Figure adapted from \cite{gg16}.} 
\label{fig:csb} 
\end{figure}

Results of recent four-body NCSM calculations of the $A$=4 hypernuclei 
\cite{gg15,gg16}, using the Bonn-J\"{u}lich LO $\chi$EFT SU(3)-based 
$YN$ interaction model \cite{LO06} with momentum cutoff in the range 
$\Lambda$=550--700~MeV, are shown in Fig.~\ref{fig:csb}. Plotted on the 
l.h.s. are the calculated $0^+_{\rm g.s.}\to 1^{+}_{\rm exc}$ excitation 
energies in \lamb{4}{H} and in \lamb{4}{He}, both of which are found to 
increase with $\Lambda$ such that somewhere between $\Lambda$=600 and 
650 MeV the $\gamma$-ray measured values of $E_{\rm x}$ are reproduced. 
The $\Lambda-\Sigma^0$ mixing CSB splitting $\Delta E_x$ obtained by 
using Eq.~(\ref{eq:OME}) also increases with $\Lambda$ such that for 
$\Lambda$=600~MeV the calculated value $\Delta E_x = \Delta B_{\Lambda}^{
\rm calc}(0^+_{\rm g.s.})-\Delta B_{\Lambda}^{\rm calc}(1^+_{\rm exc})=330
\pm 40$~keV agrees with the measured value of $E_{\rm x}$(\lamb{4}{He})$-E_{
\rm x}$(\lamb{4}{H})$=320\pm 20$~keV deduced from Fig.~\ref{fig:A=4} (right). 

Plotted on the r.h.s. of Fig.~\ref{fig:csb} is the $\hbar\omega$ dependence 
of $\Delta B^{J}_{\Lambda}$, including $V_{\rm CSB}$ from Eq.~(\ref{eq:OME}) 
and using $N_{\rm max}\to\infty$ extrapolated values for each of the four 
possible $B^{J}_{\Lambda}$ values calculated at cutoff $\Lambda$=600~MeV. 
Extrapolation uncertainties for $\Delta B^{J}_{\Lambda}$ are 10 to 20~keV. 
$\Delta B^{J=0}_{\Lambda}$ varies over the spanned $\hbar\omega$ range by 
a few keV, whereas $\Delta B^{J=1}_{\Lambda}$ varies by up to $\sim$30~keV. 

Fig.~\ref{fig:csb} demonstrates a strong (moderate) cutoff dependence of 
$\Delta B^{J=0}_{\Lambda}$ ($\Delta B^{J=1}_{\Lambda}$): 
\begin{equation} 
\Delta B^{J=0}_{\Lambda}=177^{+119}_{-147}~{\rm keV},\,\,\,\,\,\, 
\Delta B^{J=1}_{\Lambda}=-215^{+43}_{-41}~{\rm keV}. 
\label{eq:DelB} 
\end{equation} 
The opposite signs and roughly equal sizes of these $\Delta B^J_{\Lambda}$ 
values follow from the dominance of the $^1S_0$ contact term (CT) in the 
$\Lambda\Sigma$ coupling potential of the LO $\chi$EFT $YN$ Bonn-J\"{u}lich 
model~\cite{LO06}, whereas the PS SU(3)-flavor octet (${\bf 8_{\rm f}}$) 
meson-exchange contributions are relatively small and of opposite sign to 
that of the $^1S_0$ CT contribution. This paradox is resolved by noting that 
regularized pieces of Dirac $\delta(\bf r)$ potentials that are discarded in 
the classical DvH treatment survive in the LO $\chi$EFT PS meson-exchange 
potentials. Suppressing such a zero-range regulated piece of CSB OPE within 
the full LO $\chi$EFT $A$=4 hypernuclear wavefunctions gives~\cite{gg16} 
\begin{equation} 
{\rm OPE (DvH):}\,\,\,\,\,\,\Delta B^{J=0}_{\Lambda}\approx 
175\pm 40~{\rm keV},\,\,\,\,\,\, \Delta B^{J=1}_{\Lambda}\approx 
-50\pm 10~{\rm keV},  
\label{eq:OPE} 
\end{equation}  
with smaller momentum cutoff dependence uncertainties than in 
Eq.~(\ref{eq:DelB}). Both Eqs.~(\ref{eq:DelB}) and (\ref{eq:OPE}) agree 
within uncertainties with the CSB splittings $\Delta B^J_{\Lambda}$ marked 
in Fig.~\ref{fig:A=4}.

\section{CSB in $p$-shell hypernuclei}
\label{sec:p} 

Recent cluster-model work \cite{hiyama09,zhang12,hiyama12} fails to explain 
CSB splittings in $p$-shell mirror hypernuclei, apparently for disregarding 
the underlying CS $\Lambda\Sigma$ coupling potential. In the approach reviewed 
here, one introduces an effective CS $\Lambda\Sigma$ central interaction 
${\cal V}_{\Lambda\Sigma}={\bar V}_{\Lambda\Sigma}+\Delta_{\Lambda\Sigma}\,{
\vec s}_N\cdot{\vec s}_Y$, where ${\vec s}_N$ and ${\vec s}_Y$ are the nucleon 
and hyperon spin-$\frac{1}{2}$ vectors. The $p$-shell $0p_N0s_Y$ matrix 
elements ${\bar V}^{0p}_{\Lambda\Sigma}$ and $\Delta^{0p}_{\Lambda\Sigma}$, 
listed in the caption to Table~\ref{tab:pshell}, follow from the shell-model 
reproduction of hypernuclear $\gamma$-ray transition energies by 
Millener~\cite{millener12} and are smaller by roughly factor of two than the 
corresponding $s$-shell $0s_N0s_{Y}$ matrix elements, therefore resulting in 
smaller $\Sigma$ hypernuclear admixtures and implying that CSB contributions 
in the $p$ shell are weaker with respect to those in the $A=4$ hypernuclei 
also by a factor of two. To evaluate these CSB contributions, the 
single-nucleon expression (\ref{eq:OME}) is extended by summing over 
$p$-shell nucleons~\cite{gal15}:  
\begin{equation} 
V_{\rm CSB} = -0.0297\,\frac{1}{\sqrt{3}}
\sum_j{({\bar V}^{0p}_{\Lambda\Sigma}+\Delta^{0p}_{\Lambda\Sigma}\,
{\vec s}_j\cdot{\vec s}_Y)\,\tau_{jz}}. 
\label{eq:VCSB} 
\end{equation} 

Results of applying this effective $\Lambda\Sigma$ coupling model to several 
pairs of g.s. levels in $p$-shell hypernuclear isomultiplets are given in 
Table~\ref{tab:pshell}, abridged from Ref.~\cite{gal15}. All pairs except for 
$A=7$ are g.s. mirror hypernuclei identified in emulsion \cite{davis05} where 
binding energy systematic uncertainties are largely canceled out in forming 
the listed $\Delta B_{\Lambda}^{\rm exp}$ values. The $B_{\Lambda}$ data 
selected for the $A$=7 (\lamb{7}{He},~\lamb{7}{Li}$^{\ast}$,~\lamb{7}{Be}) 
isotriplet of lowest ${\frac{1}{2}}^+$ levels deserve discussion. Recall that 
the $^6$Li core state of \lamb{7}{Li}$^{\ast}$ is the $0^+$ $T$=1 at 3.56 MeV, 
whereas the core state of \lamb{7}{Li}$_{\rm g.s.}$ is the $1^+$ $T$=0 g.s. 
Thus, to obtain $B_{\Lambda}$(\lamb{7}{Li}$^{\ast}$) from $B_{\Lambda}
$(\lamb{7}{Li}$_{\rm g.s.}$) one makes use of the observation 
of a 3.88~MeV $\gamma$-ray transition \lamb{7}{Li}$^{\ast}\to
\gamma$+\lamb{7}{Li}~\cite{tamura00}. While emulsion $B_{\Lambda}^{\rm exp}
$(g.s.) values~\cite{davis05} were used for the \lamb{7}{Be}--\lamb{7}{Li}$^{
\ast}$ pair, more recent counter measurements that provide absolute energy 
calibrations relative to precise values of free-space known masses were used 
for the \lamb{7}{Li}$^{\ast}$--\lamb{7}{He} pair~\cite{botta17} (FINUDA 
for \lamb{7}{Li}$_{\rm g.s.}$ $\pi^-$ decay~\cite{botta09} and JLab for 
\lamb{7}{He} electroproduction~\cite{JLabL7He}). Note that the value reported 
by FINUDA for $B_{\Lambda}$(\lamb{7}{Li}$_{\rm g.s.}$), 5.85$\pm$0.17~MeV, 
differs from the emulsion value of 5.58$\pm$0.05~MeV. Recent $B_{\Lambda}$ 
values from JLab electroproduction experiments for \lamb{9}{Li}~\cite{JLabL9Li} 
and \lam{10}{Be}~\cite{JLabL10Be} were not used for lack of similar data on 
their mirror partners. 

\begin{table}[htb] 
\caption{$\langle V_{\rm CSB}\rangle$ contributions (in keV) to $\Delta 
B^{\rm calc}_{\Lambda}$ in $p$-shell hypernuclei g.s. isomultiplets, 
using $\Lambda\Sigma$ coupling matrix elements ${\bar V}^{0p}_{\Lambda\Sigma}
$=1.45~MeV and $\Delta^{0p}_{\Lambda\Sigma}$=3.04~MeV in Eq.~(\ref{eq:VCSB}). 
A similar calculation for the $s$-shell $A$=4 mirror hypernuclei~\cite{gal15} 
is included for comparison. Listed values of $\Delta B_{\Lambda}^{\rm exp}$ 
are based on g.s. emulsion data~\cite{davis05} except for 
\lamb{4}{He}--\lamb{4}{H} \cite{MAMI16} and \lamb{7}{Li}$^{\ast}-$\lamb{7}{He} 
\cite{botta17}.} 
\begin{center} 
\begin{tabular}{ccccccc}
\br 
\lamb{A}{Z$_{>}$}--\lamb{A}{Z$_{<}$} & \lamb{4}{He}--\lamb{4}{H} & 
\lamb{7}{Be}--\lamb{7}{Li}$^{\ast}$ & \lamb{7}{Li}$^{\ast}$--\lamb{7}{He} & 
\lamb{8}{Be}--\lamb{8}{Li} & \lamb{9}{B}--\lamb{9}{Li} & 
\lam{10}{B}--\lam{10}{Be} \\ 
$I,J^{\pi}$ & $\frac{1}{2},0^+$ & $1,{\frac{1}{2}}^+$ & $1,{\frac{1}{2}}^+$ & 
$\frac{1}{2},1^-$ & $1,{\frac{3}{2}}^+$ & $\frac{1}{2},1^-$ \\ 
\mr 
$\langle V_{\rm CSB}\rangle$ & 232 & 50 & 50 & 119 & 81 & 17 \\ 
$\Delta B_{\Lambda}^{\rm calc}$ & 226 & $-$17& $-$28 & $+$49 & $-$54 & 
$-$136 \\ 
$\Delta B_{\Lambda}^{\rm exp}$ & 233$\pm$92 & $-$100$\pm$90 & 
$-$20$\pm$230 & $+$40$\pm$60 & $-$210$\pm$220 & $-$220$\pm$250 \\ 
\br 
\end{tabular} 
\label{tab:pshell} 
\end{center} 
\end{table}

The $\langle V_{\rm CSB}\rangle$ $p$-shell entries listed in 
Table~\ref{tab:pshell} were calculated with $\Lambda$-hypernuclear 
weak-coupling shell-model wavefunctions in terms of nuclear-core g.s. leading 
SU(4) supermultiplet components, except for $A=8$ where the first excited 
nuclear-core level had to be admixed in. The listed $A=7-10$ values of 
$\langle V_{\rm CSB}\rangle$ exhibit strong SU(4) correlations, highlighted 
by the enhanced value of 119~keV for the SU(4) nucleon-hole configuration 
in \lamb{8}{Be}--\lamb{8}{Li} with respect to the modest value of 17~keV 
for the SU(4) nucleon-particle configuration in \lam{10}{B}--\lam{10}{Be}. 
This enhancement follows from the relative magnitudes of the Fermi-like 
interaction term ${\bar V}^{0p}_{\Lambda\Sigma}$ and its Gamow-Teller partner 
term $\Delta^{0p}_{\Lambda\Sigma}$. Noting that both the $A=4$ and $A=8$ 
mirror hypernuclei correspond to SU(4) nucleon-hole configuration, the roughly 
factor two ratio of $\langle V_{\rm CSB}{\rangle}_{A=4}$=232~keV to $\langle 
V_{\rm CSB}{\rangle}_{A=8}$=119~keV reflects the approximate factor of two 
for $0s_N0s_Y$ to $0p_N0s_Y$ $\Lambda\Sigma$ matrix elements discussed above. 
However, in distinction from the $A$=4 g.s. isodoublet where $\Delta B_{
\Lambda}\approx \langle V_{\rm CSB}\rangle$, the increasingly negative Coulomb 
contributions in the $p$-shell overcome the positive $\langle V_{\rm CSB}{
\rangle}$ contributions, with $\Delta B_{\Lambda}$ becoming negative definite 
for $A\geq 9$. 

Comparing $\Delta B_{\Lambda}^{\rm calc}$ with $\Delta B_{\Lambda}^{\rm exp}$ 
in Table~\ref{tab:pshell}, we note the reasonable agreement reached between 
the $\Lambda\Sigma$ coupling model calculation and experiment for all five 
pairs of $p$-shell hypernuclei listed here. Extrapolating to heavier 
hypernuclei, one might naively expect negative values of $\Delta B_{\Lambda}^{
\rm calc}$. However, this assumes that the negative Coulomb contribution 
remains as large upon increasing $A$ as it is in the beginning of the $p$ 
shell, which need not be the case. As nuclear cores beyond $A=9$ become more 
tightly bound, the $\Lambda$ hyperon is unlikely to compress these nuclear 
cores as much as it does in lighter hypernuclei, so that the additional 
Coulomb repulsion in \lam{12}{C}, for example, over that in \lam{12}{B} may 
not be sufficiently large to offset the attractive CSB contribution to 
$B_{\Lambda}$(\lam{12}{C})$-B_{\Lambda}$(\lam{12}{B}), in agreement with the 
value 50$\pm$110~keV suggested recently for this $A$=12 $B_{\Lambda}$(g.s.) 
splitting using FINUDA and JLab counter measurements~\cite{botta17}. 
In making this argument one relies on the expectation, based on SU(4) 
supermultiplet fragmentation patterns in the $p$ shell, that $\langle 
V_{\rm CSB}\rangle$ does not exceed $\sim$100~keV. 

Some implications of the state dependence of CSB splittings, e.g. the large 
difference between the calculated $\Delta B_{\Lambda}(0^+_{\rm g.s.})$ and 
$\Delta B_{\Lambda}(1^+_{\rm exc})$ in the $s$ shell, Eqs.~(\ref{eq:DelB}) or 
(\ref{eq:OPE}), are worth noting also in the $p$ shell. The most spectacular 
one concerns the \lam{10}{B} g.s. doublet splitting, where adding the $\Lambda
\Sigma$ coupling model CSB contribution of $\approx -27$~keV to the 
$\approx$110~keV CS $1^-_{\rm g.s.}\to 2^-_{\rm exc}$ g.s. doublet excitation 
energy calculated in this model~\cite{millener12} helps bring it down well 
below 100~keV, which is the upper limit placed on it from past searches for 
a $2^-_{\rm exc}\to 1^-_{\rm g.s.}$ $\gamma$-ray 
transition~\cite{chrien90,tamura05}.

\section{Summary and Outlook} 
\label{sec:sum} 

The recent J-PARC E13-experiment observation of a 1.41 MeV \lamb{4}{He}($1^+_{
\rm exc}\to 0^+_{\rm g.s.}$) $\gamma$-ray transition \cite{E13}, and the 
recent MAMI-A1 determination of $B_{\Lambda}$(\lamb{4}{H}) to better than 100 
keV \cite{MAMI15,MAMI16}, plus the recenly approved J-PARC E63 experiment to 
remeasure the \lamb{4}{H}($1^+_{\rm exc}\to 0^+_{\rm g.s.}$) $\gamma$-ray 
transition, arose renewed interest in the sizable CSB already confirmed 
thereby in the $A$=4 mirror hypernuclei. It was shown in the present report 
how a relatively large $\Delta B_{\Lambda}(0^+_{\rm g.s.})$ CSB contribution 
of order 250~keV, in rough agreement with experiment, arises in ab-initio 
four-body calculations \cite{gg15,gg16} using $\chi$EFT $YN$ interactions 
already at LO. 

In $p$-shell hypernuclei, a $\Lambda\Sigma$ coupling shell-model approach 
was shown to reproduce CSB splittings of g.s. binding energies~\cite{gal15}. 
More theoretical work in this mass range, and beyond, is needed to understand 
further and better the salient features of $\Lambda\Sigma$ dynamics 
\cite{galmil13}. On the experimental side, the recently proposed ($\pi^-,K^0$) 
reaction \cite{agnello16} should be explored, in addition to the standard 
($\pi^+,K^+$) reaction, in order to study simultaneously two members of 
a given $\Lambda$ hypernuclear isomultiplet, for example reaching both 
\lam{12}{B} and \lam{12}{C} on a carbon target.

\ack 
A.G. acknowledges instructive discussions with John Millener, as well as the 
gracious hospitality extended by the organizers of SPRING 2017 at Ischia, 
Italy. The research of D.G. was supported by the Grant Agency of the Czech 
Republic (GACR), Grant No. P203/15/04301S.

\end{document}